\documentclass[sigconf]{acmart}
\AtBeginDocument{%
  }

\setcopyright{acmlicensed}
\copyrightyear{2018}
\acmYear{2018}
\acmDOI{XXXXXXX.XXXXXXX}
\acmConference[Conference acronym 'XX]{Make sure to enter the correct
  conference title from your rights confirmation email}{June 03--05,
  2018}{Woodstock, NY}
\acmISBN{978-1-4503-XXXX-X/2018/06}

\usepackage{multirow}
\usepackage{booktabs} 
\usepackage{arydshln} 
\usepackage{makecell} 
\usepackage[table]{xcolor} 
\usepackage{algorithm}
\usepackage{algorithmicx}
\usepackage{algpseudocode}
\usepackage{float}

\begin{document}

\title{An Efficient Framework for Whole-Page Reranking via Single-Modal Supervision}

\author{Zishuai Zhang}
\affiliation{%
  \institution{School of Artifical Intelligence, Beihang University}
  \city{Beijing}
  \country{China}}
\email{zhangzishuai@buaa.edu.cn}
  
\author{Sihao Yu}
\authornote{Corresponding author}
\affiliation{%
  \institution{Baidu Inc.}
  \city{Beijing}
  \country{China}}
\email{yush93@qq.com}

\author{Wenyi Xie}
\affiliation{%
  \institution{Baidu Inc.}
  \city{Beijing}
  \country{China}}
\email{xiewenyi@baidu.com}

\author{Ying Nie}
\affiliation{%
  \institution{Baidu Inc.}
  \city{Beijing}
  \country{China}}
\email{niezengyings@gmail.com}

\author{Junfeng Wang}
\affiliation{%
  \institution{Baidu Inc.}
  \city{Beijing}
  \country{China}}
\email{junfengzju@163.com}

\author{Zhiming Zheng}
\affiliation{%
  \institution{School of Artifical Intelligence, Beihang University}
  \city{Beijing}
  \country{China}}
\email{zhengzhiming0130@163.com}

\author{Dawei Yin}
\affiliation{%
  \institution{Baidu Inc.}
  \city{Beijing}
  \country{China}}
\email{yindawei@acm.org}  

\author{Hainan Zhang}
\authornote{Corresponding author}
\affiliation{%
  \institution{School of Artifical Intelligence, Beihang University}
  \city{Beijing}
  \country{China}}
\email{zhanghainan@buaa.edu.cn}

\renewcommand{\shortauthors}{Zhang et al.}

\begin{abstract}
The whole-page reranking plays a critical role in shaping the user experience of search engines, which integrates retrieval results from multiple modalities, such as documents, images, videos, and LLM outputs. Existing methods mainly rely on large-scale human-annotated data, which is costly to obtain and time-consuming.
This is because whole-page annotation is far more complex than single-modal: it requires assessing the entire result page while accounting for cross-modal relevance differences. Thus, how to improve whole-page reranking performance while reducing annotation costs is still a key challenge in optimizing search engine result pages(SERP).
In this paper, we propose SMAR, a novel whole-page reranking framework that leverages strong Single-modal rankers to guide Modal-wise relevance Alignment for effective Reranking, using only limited whole-page annotation to outperform fully-annotated reranking models.
Specifically, high-quality single-modal rankers are first trained on data specific to their respective modalities. Then, for each query, we select a subset of their outputs to construct candidate pages and perform human annotation at the page level. Finally, we train the whole-page reranker using these limited annotations and enforcing consistency with single-modal preferences to maintain ranking quality within each modality.
Experiments on the Qilin and Baidu datasets demonstrate that SMAR reduces annotation costs by about 70-90\% while achieving significant ranking improvements compared to baselines. Further offline and online A/B testing on Baidu APPs also shows notable gains in standard ranking metrics as well as user experience indicators, fully validating the effectiveness and practical value of our approach in real-world search scenarios.
\end{abstract}



\maketitle

\section{Introduction}
\begin{figure}[t]
    \centering
    \includegraphics[width=0.75\columnwidth]{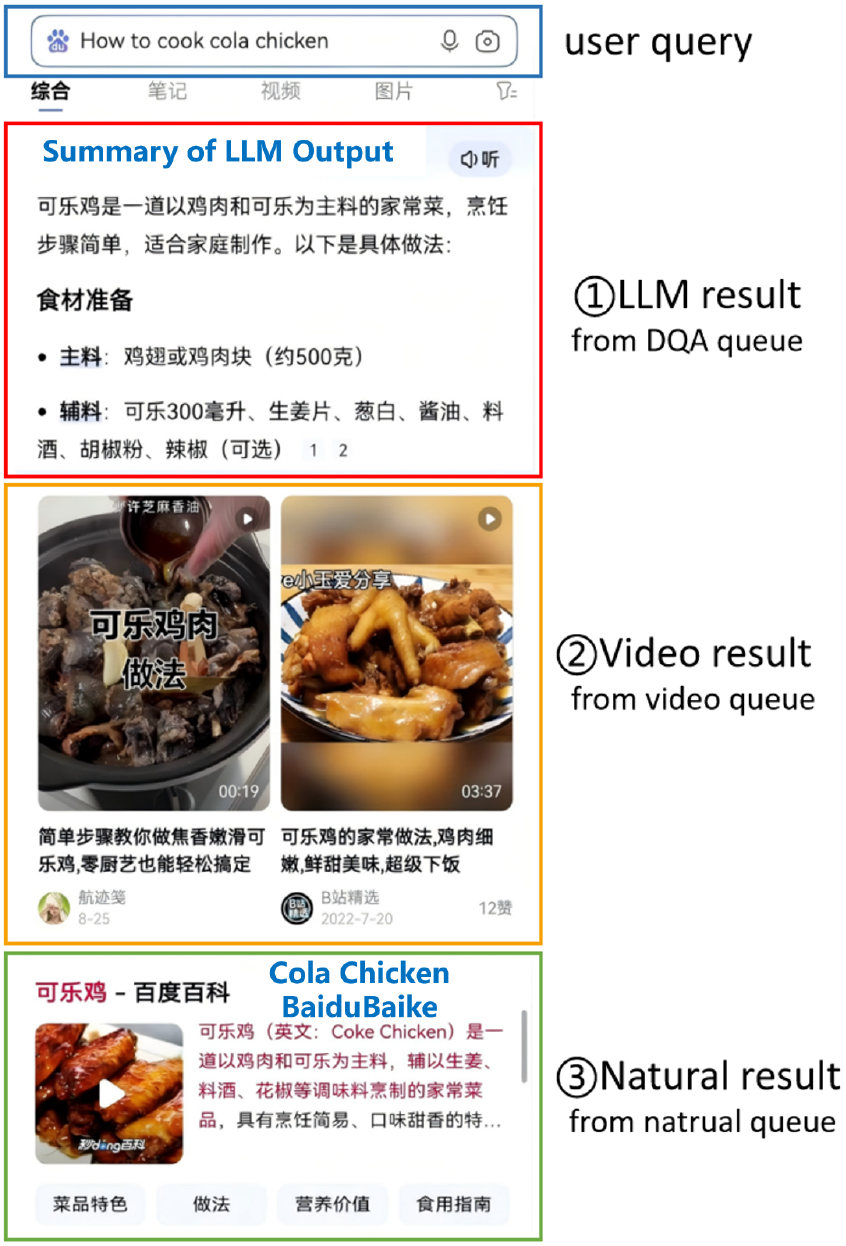}
    \caption{An example of heterogeneous retrieval and reranking on a mobile search interface. Given a user query ``How to cook cola chicken'', candidate results are recalled from multiple modality-specific queues, including LLM outputs, video content, and natural text. }
    \Description{The figure illustrates an example of heterogeneous retrieval on a mobile search interface. 
A user query “How to cook cola chicken” triggers three types of results: 
(1) a large language model output providing a textual recipe summary, 
(2) video results from the video queue showing cooking demonstrations, 
and (3) natural text results from the natural queue such as encyclopedia entries. 
The example demonstrates how multimodal candidates from different sources are jointly reranked and displayed to the user.}
    \label{fig:baiduAPP}
\end{figure}

Search engines have evolved from simple document retrieval systems into complex platforms that must seamlessly integrate information across multiple modalities, including text, images, videos, and large language model(LLM) outputs~\cite{wang2016beyond,zhang2018relevance}. A central component of this process is whole-page reranking~\cite{mao2024whole}, which determines the final ordering and presentation of heterogeneous results on the search engine results page (SERP). As shown in Figure~\ref{fig:baiduAPP}, the SERP contains various components corresponding to the user's query. Unlike single-modal ranking, whole-page reranking requires balancing diverse signals of relevance, consistency, and user satisfaction. Therefore, whole-page reranking is both crucial for user experience and technically challenging for SERP. 

Existing methods~\cite{li2023pretrained, li2025fultr, mao2024whole} for whole-page reranking in search engines typically rely on large-scale human-annotated data. Unlike labeling relevance for individual documents or images, whole-page annotations must evaluate global ranking quality while also accounting for cross-modal relevance differences. Given a user query, annotators not only need to determine the relevance order within the same modality but also assess the relative relevance across different modalities. For example, when a user searches for "How to make Coke Chicken Wings?", it is generally better to rank video results with detailed step-by-step demonstrations higher than dry text descriptions. Therefore, page-level annotations are far more costly and time-consuming than single-modal annotations. 


\begin{figure}[!t]
    \centering
    \includegraphics[width=0.7\columnwidth]{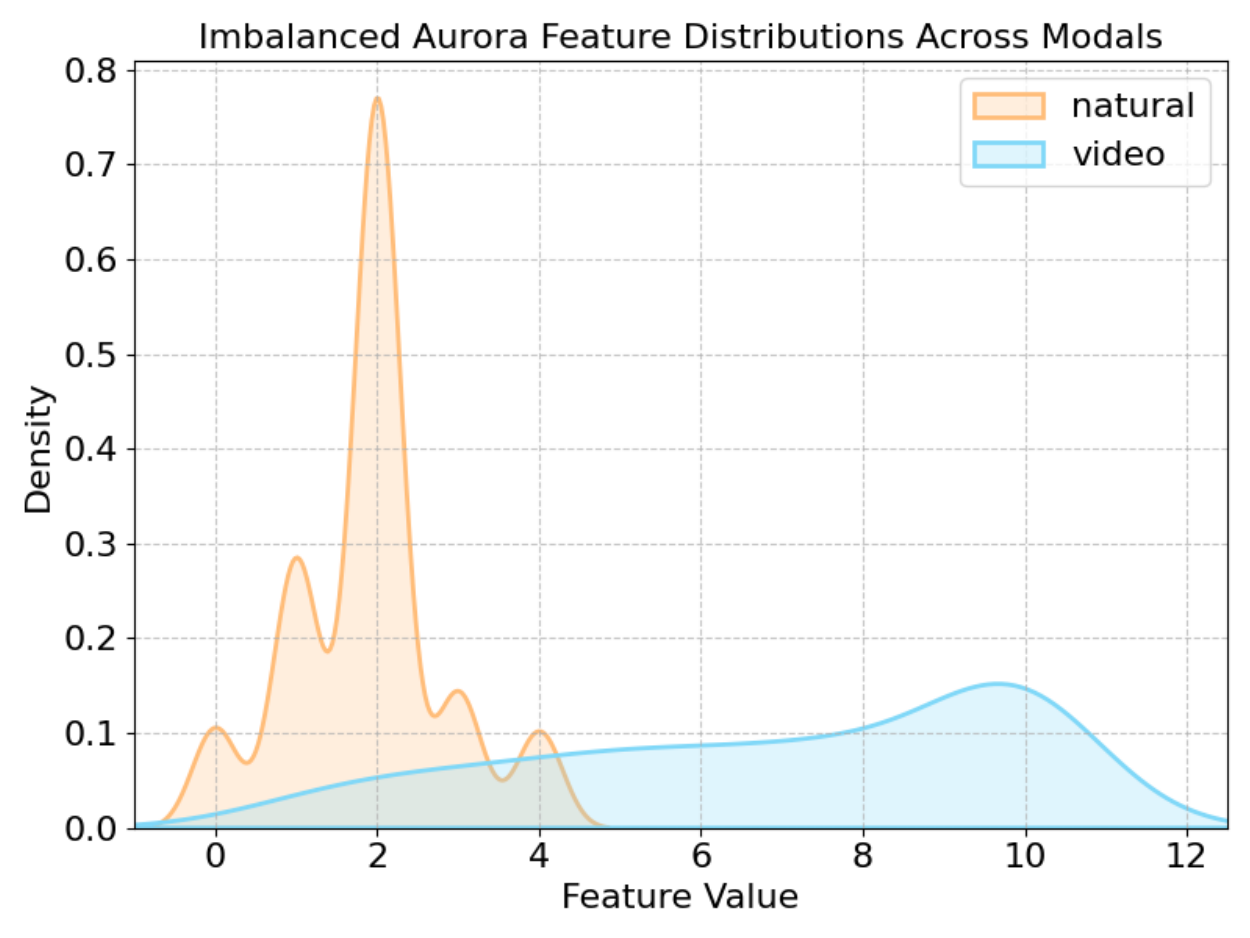}
    \caption{Imbalanced Upstream Model Scores Distributions Across Modals. Upstreammodel scores is a common feature score computed by the Single-Modal upstream models in Baidu search engine.}
    \Description{The figure shows the distribution of Upstream Model Scores across two modalities: natural text and video. 
    The natural modality exhibits a sharp peak around lower feature values, indicating a concentrated distribution, 
    whereas the video modality shows a broader and higher-value distribution. 
    The visualization highlights the imbalance in upstream model scores distributions between modalities.}
    \label{fig:Aurora_distribution}
\end{figure}

\begin{figure}[!t]
    \centering
    \includegraphics[width=0.9\columnwidth]{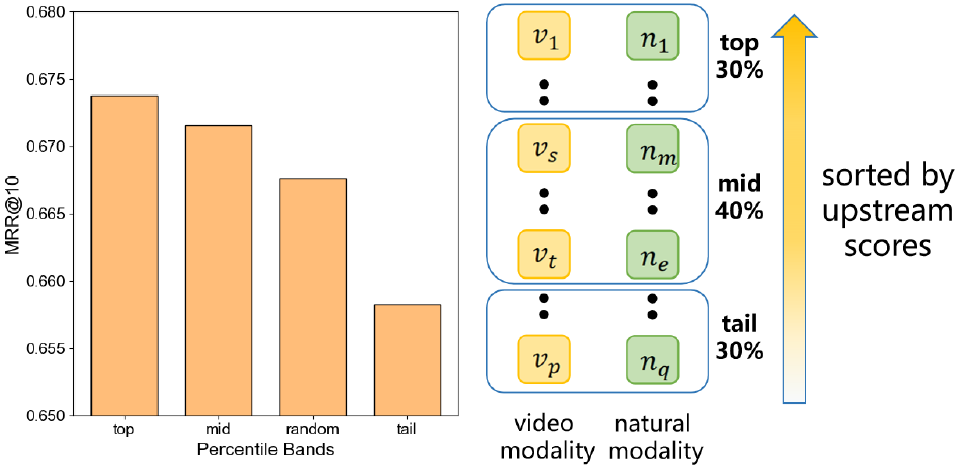}
    \caption{Performance of Percentile bands on Qilin. The video queue and the natural result queue are sorted by upstream model scores. ``Random'' denotes randomly selecting 30\% of the data from each of the two queues.}
    \Description{
        This figure illustrates the impact of selecting different percentile bands on retrieval performance (MRR@10) for a multimodal system named Qilin. 
        On the left, a bar chart shows that selecting the top 30\% of samples by upstream model scores achieves the highest MRR@10 (approximately 0.6738), followed by 30-70\% (0.6715), random 30\% (0.6676), and the tail 70-100\% (0.6582). 
        On the right, a schematic diagram explains how data from video and natural language modalities are sorted by upstream scores into three bands: top 30\%, mid 40\%, and tail 30\%. 
        Arrows indicate that the top-performing band corresponds to high-scoring samples from both modalities, while lower bands contain progressively less confident predictions. 
        The visual emphasizes that prioritizing high-confidence samples improves overall retrieval accuracy.
    }
    \label{fig:motivation}
\end{figure}

\begin{figure*}[!t]
    \centering
    \includegraphics[width=0.85\linewidth]{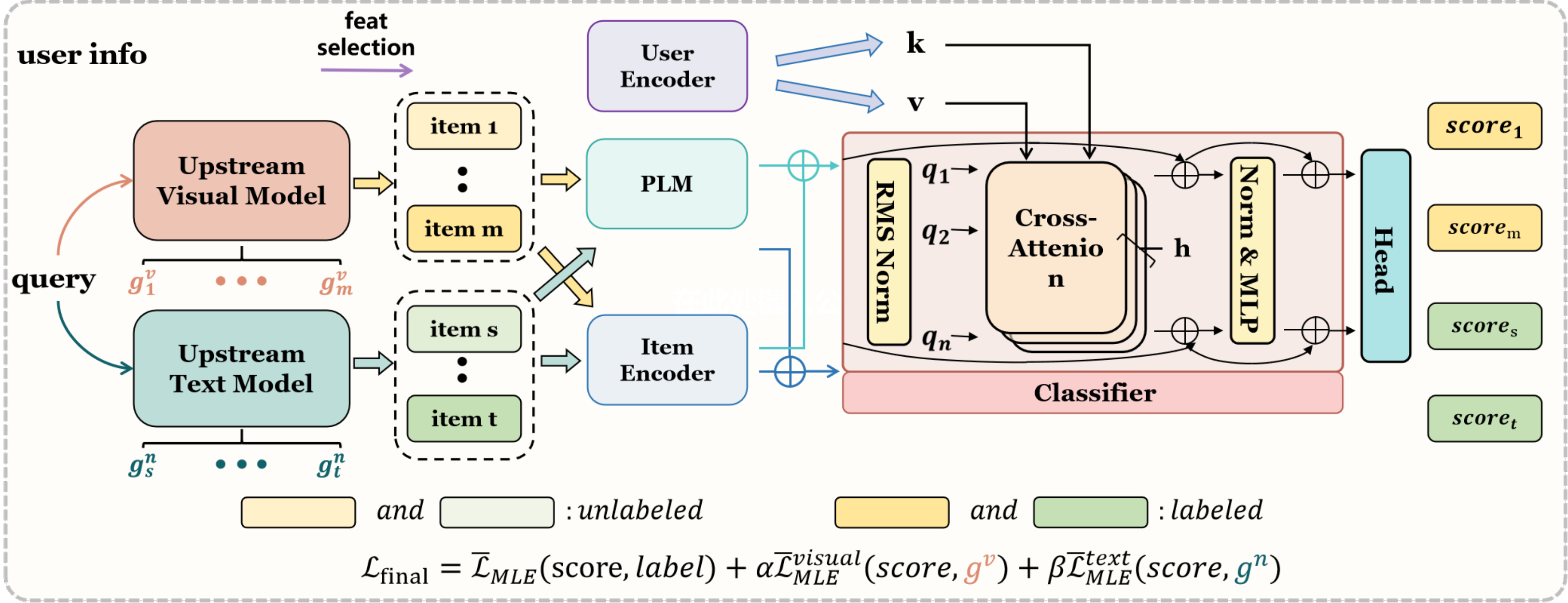}
    \caption{Overall architecture of the proposed user–item interaction model.
    \textcircled{{\small 1}}The encoder integrates query semantics with item representations, where a pretrained language model (PLM) encodes query–item textual signals and an item feature extractor incorporates auxiliary item information. 
    \textcircled{{\small 2}}User-specific features are independently processed by a user feature extractor to produce user embeddings, which are injected into the classifier as keys and values. 
    \textcircled{{\small 3}}The classifier employs cross-attention to capture fine-grained interactions between user embeddings and item representations, followed by normalization and multi-layer perceptron (MLP) modules. The head outputs the final relevance scores for candidate items.
   }
    \Description{A schematic of the user-item interaction architecture. User information is processed by a user encoder to produce a user embedding that serves as keys and values for the classifier. For each candidate item paired with a query, an encoder combines a pretrained language model (for textual signals) and an item feature extractor to form item embeddings. The classifier applies cross-attention between the user embedding and the item embeddings, followed by normalization and an MLP. A final head outputs a relevance score for each item, yielding a ranked list.}
    \label{fig:overall architecture}
\end{figure*}

Meanwhile, strong single-modal rankers achieve remarkable performance within their respective domains, such as BERT-based text models~\cite{devlin2019bert, qwen3embedding} and vision-language models~\cite{Qwen2-VL}. However, naively extending these models to whole-page reranking is nontrivial. Simple fusion methods~\cite{radford2021learning} fail to capture interdependencies among modalities, while naive combinations~\cite{zhang2025notellm} often lead to suboptimal rankings due to mismatched relevance distributions across modalities. This is because candidates originate from different modalities with distinct schemas and presentation forms, leading to large discrepancies in feature distributions, structural forms, and numeric scales across sources. Figure~\ref{fig:Aurora_distribution} shows that upstream single-modality features for video and natural text results follow markedly different distributions, highlighting the lack of cross-modal comparability. This heterogeneity makes a single global model hard to calibrate under distribution shifts, inflates engineering and labeling overhead, and ultimately constrains both the attainable gains of cross-modality ranking models and the end-user experience on the SERP. 

Thus, the key research question becomes: \textbf{How can we exploit high-quality single-modal rankers to enhance whole-page reranking while minimizing reliance on costly whole-page annotations?} To this end, we conduct a careful analysis of the annotation quality of single-modal rankers on whole-page reranking data. Specifically, we evaluate the performance of a multimodal reranking model on the Qilin dataset using four groups:  a randomly selected 30\%, the top 30\%, middle 40\%, and tail 30\% from single-modal rankers, as shown in Figure~\ref{fig:motivation}. The results demonstrate that the higher the correlation of the single-modal ranker (i.e., the top group), the better the final multimodal reranking model performs.

In this paper, we propose SMAR, a novel whole-page reranking framework that leverages strong Single-modal rankers to guide Modal-wise relevance Alignment for effective Reranking. Instead of depending solely on costly full-page annotations, SMAR introduces a cross-modal distillation mechanism where single-modal experts provide fine-grained relevance signals to guide the whole-page reranking model. Specifically, high-quality single-modal rankers are first trained on their modality-specific data. Then, we propose two budget-aware annotation strategies to form candidate pages and perform annotations at the page level: the Top-P strategy and the Iso-label anchors strategy. Top-p annotates only the highest-scoring fraction within each modality while leaving the remainder to weak supervision from upstream scores. Iso-label anchors annotate cross-modal candidates at comparable upstream scores to align preference scales and propagate supervision to the whole-page reranker. Finally, the whole-page reranker is trained using these limited annotations while enforcing consistency with single-modal preferences to preserve intra-modality ranking quality.

Extensive experiments on the Qilin and Baidu datasets demonstrate that SMAR reduces annotation costs by approximately 70-90\% while achieving significant improvements over fully annotated reranking baselines. Moreover, both offline evaluations and online A/B testing in Baidu Apps validate the practical effectiveness of SMAR, yielding 0.86\% and 0.25\% gains in NDCG and CTR as ranking metrics, as well as 1.58\% and 0.33\% gains in $\Delta GSB$ and next-day retained users. These results highlight the promise of aligning single-modal expertise with limited whole-page supervision as a scalable solution for improving SERP quality in real-world search scenarios. ~\footnote{Our code and data are available at https://github.com/zzs97str/SMAR.} Our contributions can be summarized as follows:
\begin{itemize}
    \item We highlight the annotation bottleneck in whole-page reranking and motivate the need for methods that leverage single-modal rankers to reduce costs.
    \item We propose two novel modal-wise relevance alignment strategies that distill strong single-modal signals into whole-page reranking by preserving intra-modality preference orderings.
    \item We conduct extensive experiments on large-scale industrial datasets and real-world platforms, demonstrating that SMAR achieves both annotation efficiency and practical ranking improvements. We also publish an industrial corpus for whole-page reranking research.
\end{itemize}

\section{Model}

\subsection{Preliminaries}
In practical Search Engine systems, candidate items are often retrieved from multiple heterogeneous sources, such as text corpora, multimedia collections, or structured databases. Each source provides a distinct set of candidate items with potentially different feature spaces and relevance distributions. Let $\mathcal{U}$ denote the user space, $\mathcal{Q}$ the query space, and $\{\mathcal{I}^{1}, \mathcal{I}^{2}, \dots, \mathcal{I}^{M}\}$ the candidate item sets retrieved from $M$ heterogeneous sources. Given a user $u \in \mathcal{U}$ and a query $q \in \mathcal{Q}$, the system first performs multimodal retrieval to construct a personalized unified candidate set
\[
\mathcal{I} = \bigcup_{m=1}^{M} \mathcal{I}^{m}.
\]

The goal of personalized heterogeneous ranking is to learn a multimodal scoring function 
\[
f: \mathcal{U} \times \mathcal{Q} \times \mathcal{I} \rightarrow \mathbb{R},
\]
that assigns each item $i \in \mathcal{I}$ a relevance score $s_i = f(u, q, i)$, such that items from different modals can be compared in a shared semantic space and ranked in descending order of relevance:
\[
\pi(u, q) = \operatorname*{argsort}_{i \in \mathcal{I}} \; s_i
\]
where $\pi(u, q)$ denotes the final ranked list.
Although this formulation is conceptually straightforward, learning an accurate scoring function $f(u,q,i)$ is highly non-trivial. The joint space $\mathcal{U}\times\mathcal{Q}\times\mathcal{I}$ is extremely large and high-dimensional. Training a reliable model over this space requires massive amounts of human-labeled data, which is expensive and leads to low efficiency in practice.

\subsection{Overall Framework}

We leverage upstream single-modal rankers’ fine-grained scoring signals for heterogeneous reranking, avoiding large-scale manual annotation. Specifically, the scores assigned to items within a single-modal sequence are transformed into supervision for the multimodal reranking model through pairwise or listwise objectives. This formulation allows us to reuse the strengths of existing rankers, significantly reducing annotation cost while maintaining strong performance in heterogeneous ranking. The overall framework is shown in Figure~\ref{fig:overall architecture}.

\subsection{Single-modal Ranker Supervison}
Let $\pi^{m}(q)=\{ (i, g^{m}_i(q)) | i \in \mathcal{I}^m\}$ denote the items and their scores assigned by the upstream single-modal ranker $g^{m}$, where $g^{m}_i(q)$ is the score of item $i \in \mathcal{I}^{(m)}$ under query $q$. These scores serve as supervision signals for training the multimodal reranker $f(u,q,i)$ under two proposed supervision schemes:

\textbf{Pairwise supervision.} Given a query, for any two items $i, j \in \mathcal{I}^{(m)}$ from the same source, we define the pairwise preference as:
\[
y_{ij} = \mathbb{I}\big[g^{m}_i > g^{m}_j\big],
\]
where $\mathbb{I}[\cdot]$ is the indicator function. 

The reranker is trained to preserve this preference by minimizing a pairwise loss:
\[
\mathcal{L}_m 
= y_{ij}\cdot\max\!\big(0,\, \gamma - \big(f(u,q,i) - f(u,q,j)\big)\big),
\]
where $\gamma>0$ is the margin hyperparameter. 

The total multimodal reranker loss is
\[
\mathcal{L}_{pair} =\sum_{m \in M} \beta_m \cdot\mathcal{L}_m,
\]
where $\beta_m$ is a coefficient of modality m.

\textbf{Listwise supervision.} Alternatively, we can use listwise objectives to train the multimodal reranker.
Given a single-modal candidate list $\mathcal{I}^{m}$ with upstream scores $\{g^{m}_i\}_{i \in \mathcal{I}^{m}}$, we normalize them into a probability distribution:
\[
p^{m}(i \mid q) = \frac{\exp(g^{m}_i)}{\sum_{j \in \mathcal{I}^{m}} \exp(g^{m}_j)}.
\]
The reranker then produces a distribution over these candidates:
\[
\hat{p}(i \mid u,q) = \frac{\exp(f(u,q,i))}{\sum_{j \in \mathcal{I}^m} \exp(f(u,q,j))}.
\]
The listwise objective minimizes the discrepancy between upstream supervision and the reranker's distribution, e.g., via KL divergence:
\[
\mathcal{L}_{\text{list}} = D_{\text{KL}}\!\left(p^{m}(\cdot \mid q) \,\|\, \hat{p}(\cdot \mid u,q)\right).
\]
If only the ranking order from the upstream model is available while the exact item scores are unknown, one can adopt ListMLE~\cite{xia2008listwise}, which directly models the permutation likelihood of a ranked list. Given a permutation $\pi$ sorted by the upstream scores, the ListMLE loss is defined as:
\[
\mathcal{L}_{\text{ListMLE}} = - \sum_{k=1}^{|\mathcal{I}^{(m)}|} 
\log \frac{\exp(f(u,q,\pi_k))}{\sum_{j=k}^{|\mathcal{I}^{(m)}|} \exp(f(u,q,\pi_j))}.
\]

Importantly, the proposed supervision relies solely on relative preferences—that is, whether item A should be ranked above item B—rather than on their absolute scores. For example, the upstream model may only return binned features rather than explicit model prediction scores. In such cases, the multimodal ranking model can still be supervised using pairwise comparisons constructed between items from different bins within a single modality. This design aligns with real-world industrial settings, where information silos across teams or organizations often limit access to raw data.

\subsection{Annotation Strategy}
Solely training the multimodal reranker models on single-modal data is insufficient. In single-modal ranking, the upstream models, given a query, only consider the relevance between the content of the current modal data and the query. They do not take into account the user's specific modal preferences and which data source is most suitable for the current query. Therefore, effective heterogeneous ranking requires high-quality cross-modal annotations.
However, obtaining such high-quality and multimodal annotations is extremely costly. Manual annotation requires substantial human resources, time, and effort, especially when dealing with large-scale and diverse datasets. 
Instead of annotating the entire dataset, we focus on specific samples that are more representative or have a greater impact on the model's performance. We investigate two annotation strategies: the Top-P strategy and the binary search for iso-label anchors strategy.
\subsubsection{Top-P Strategy}
Users focus their attention on the first few results and expect to find a satisfactory item among the top entries with minimal effort~\cite{chuklin2022click}. Guided by this insight, we adopt a budget-aware annotation policy that prioritizes the highest-exposure region of each source list. 

Let $\mathcal{M}$ denote the set of modals. For query $q$ and modal $m\in\mathcal{M}$, an upstream ranker $g_m$ produces a scored list $\mathcal{C}_{m}=\{(x_{j}, g_m(x_{j}))\}_{j=1}^{n_{m}}$.
Given a budget ratio $p\in(0,1]$, we form the labeled pool by taking the top $ p\cdot n_{m}$ items from each $\mathcal{C}_{m}$.
The remaining items are left unlabeled, supervised by their upstream ranker. By selectively annotating these samples, we can improve the model's ability to handle cross-list ranking while keeping the annotation costs under control. This approach allows us to make the most of our limited resources and effectively enhance the performance of the multimodal reranker models in heterogeneous ranking scenarios. 

\subsubsection{Binary Search for Iso-Label Anchors}
\begin{figure}[!t]
    \centering
    \includegraphics[width=0.85\columnwidth]{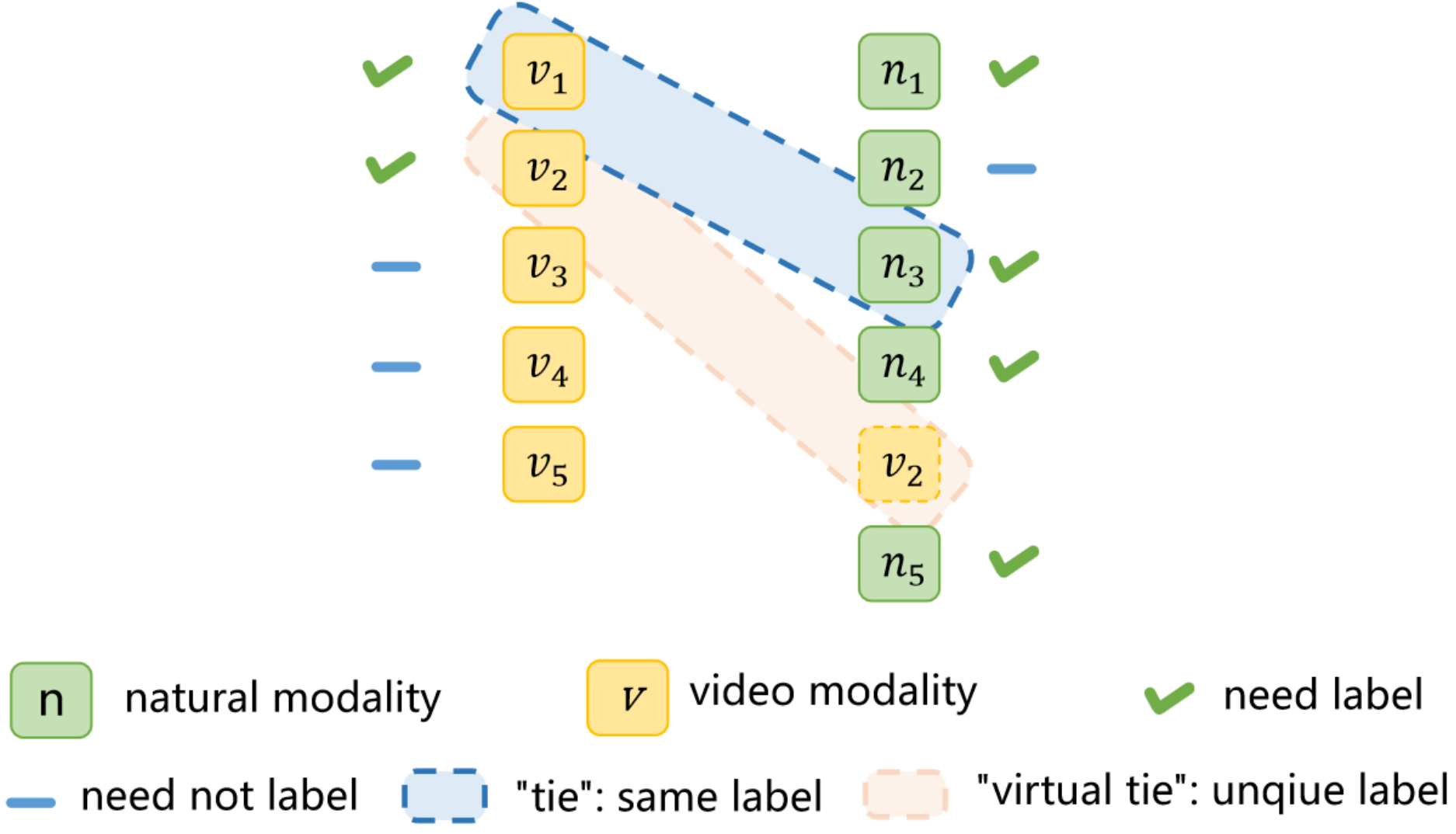}
    \caption{An example of binary search for iso-label anchors. Blue dashed boxes represent tie segments where items share the same label, and orange dashed boxes indicate virtual ties where no item in the natural queue has the same label as $v_2$, but natural items with both higher and lower labels exist, leading to the insertion of $v_2$ as a virtual tie.}
    \Description{The figure illustrates an example of binary search for iso-label anchors between two modality queues. Items from the video modality (v) and the natural modality (n) are aligned according to their labels. Blue dashed boxes represent tie segments where items share the same label, and orange dashed boxes indicate virtual ties where no item in the natural queue below the previous tie has the same label as $v_2$, but items with both higher and lower labels exist, leading to the insertion of $v_2$ as a virtual tie. Green check marks denote items that require labeling, while blue dashes mark those that do not. The example demonstrates how multimodal alignment is established by identifying pairs of items with identical labels.}

    \label{fig:binary search}
\end{figure}
Although the feature distributions of different modalities vary for the same query, when the scores produced by upstream single-modality models accurately capture modality-specific preferences, it becomes possible to establish items sharing identical or nearly identical prior satisfaction scores across modalities (``tie'') as bridges for alignment. To efficiently identify items with identical prior satisfaction scores while relying on limited labeled data, we employ a Binary Search for Iso-Label Anchors strategy. 

Formally, let $Q_1$ and $Q_2$ denote two ranked queues, where each queue is sorted by the stable upstream scores. For every $i \in Q_1$, we first label $j \in Q_2$ to check if $label(Q_1[i]) = label(Q_2[j])$. If their labels are equal, a ``tie'' is obtained. Else, we perform a binary search over $Q_2$ to locate an index $j$ such that $label(Q_1[i]) = label(Q_2[j])$. Upon a match, we obtain an equal-label anchor and move on to the next $i' \in Q_1$ to repeat the process until all items are considered in one of the two queues. If all labels in $Q_2$ are different from $label(Q_1[i'])$, we insert $i'$ into $Q_2$ to obtain a ``virtual'' tie. T denotes the total number of times we search for the ties. The larger the T, the more data we choose. For instance, after searching for 2 times, the results are shown in Figure ~\ref{fig:binary search}. For labeled items, a pointwise loss is obtained. For $n_1$ and $v_1$, a pairwise loss is obtained, etc. Pseudo-code is provided in Appendix~\ref{sec: Binary Search for Iso-Label Anchors}. Leveraging these anchors, SMAR aligns multimodal label spaces with limited supervision and distills intra-modality preferences from upstream models, enabling effective learning of score distributions across heterogeneous modalities. 

\subsection{Whole-Page Reranker}
A recent study~\cite{pobrotyn2020context} shows self-attention effectively models inter-document importance for context-aware ranking. Yet such methods neglect a key factor in heterogeneous ranking: the user. Preferences and locale shape relevance. For instance,“Java” may mean the Indonesian island for local users but the programming language for many developers, so rerankers should incorporate user features. To address this limitation, our framework introduces a cross-attention module between user embeddings and item embeddings. 

As illustrated in Figure~\ref{fig:overall architecture}, the encoder first generates semantic representations of queries and items $\mathbf{e}^{\text{sem}}_{i}$ using a pretrained language model (PLM). Inspired by \cite{zhang2025notellm}, we utilize a hybrid fusion to enhance the contribution of $e^{visual}$ to the final item embedding. 
Specifically, we first obtain the text embedding $e^{text}$ from the last token of the final layer output of the VLM, which serves as the comprehensive representation of the textual content.
Early fusion maps visual tokens into the VLM’s text space to produce an image-only visual note $e_i^{visual}$.  In late fusion, we employ a learnable gating mechanism to adaptively combine the visual representation $e_i^{visual}$ with the text representation $e^{text}$, producing the final multimodal embedding:
\begin{equation}
    z=\sigma(W[e^{visual},e^{text}]+b),
\end{equation}
\begin{equation}
    e^{\text{sem}}_{i}=z\odot e^{visual}+(1-z)\odot e^{text}.
\end{equation}
Here, $\sigma$ is the sigmoid function, $[\cdot,\cdot]$ denotes vector concatenation, and $\odot$ is the element-wise product.
In parallel, an item feature extractor encodes the numerical and categorical attributes of candidate items into high-dimensional embeddings: 
\[
\mathbf{e}^{\text{feat}}_{i} = \text{FFN}(x_{i}),
\]
where $x_{i}$ represents the structured features of item $i$. Then  semantic representations $\mathbf{e}^{\text{sem}}_{i}$ and item feature embedding $\mathbf{e}^{\text{feat}}_{i}$ are concatenated into $e_i$. 
\[
\mathbf{e}_{i} = [\mathbf{e}^{\text{sem}}_{i}; \mathbf{e}^{\text{feat}}_{i}],
\]
Likewise, a user feature extractor produces user embeddings:
\[
\mathbf{e}^{\text{user}} = \text{FFN}(x_{u}),
\]
with $x_{u}$ denoting the user features. Both user features $x_{u}$ and item features $x_{i}$ are bucketized and converted into binary form~\cite{pan2024ads}, resulting in a feature vector.

The enriched representations $\mathbf{h}_{i}$ are processed by a stack of blocks, including normalization, cross-attention mechanisms, multi-layer perceptrons (MLP), and residual connections~\cite{he2016deep}.

\begin{align}
\mathbf{h}_{i} &= \text{Multi-head CrossAttn}(Q,K,V) \\
               &= \text{Concat}(\text{head}_1, \ldots, \text{head}_H) W^O,
\end{align}
where each head is computed as
\[
\text{head}_j = \text{CrossAttn}\!\left(W_j^{Q}\mathbf{e}_{i}, \, W_j^{K}\mathbf{e}^{\text{user}}, \, W_j^{V}\mathbf{e}^{\text{user}}\right).
\]
$W_Q$, $W_K$, $W_V$ and $W^O$ are learnable projection matrices. $H$ denotes the number of attention heads, which allows the model to jointly capture multiple types of user–item interaction patterns in different subspaces.
Finally, the output head produces final ranking scores:
\[
s_{i} = \text{Head}(\text{MLP}(\mathbf{h}_{i})).
\]
This design enables each candidate item to selectively attend to relevant user signals, thereby yielding interaction-aware representations that jointly capture item relevance and user preference.

During training, we adopt a listwise objective because the Qilin dataset provides only implicit click feedback per item (clicked vs. unclicked) without graded query–item relevance labels. Concretely, we optimize the ListMLE loss. Given model scores $x_q={f_\theta(x_{qj}) }_{j=1}^{n_q}$ for query q with $n_q$ candidates and a click-derived ordering $y_q$, following Xia et al.~\cite{xia2008listwise}:
\begin{align}
\mathcal{L}_{\text{MLE}}(x_q,y_q) &= - \log \big( P(y_q \mid x_q; f) \big), \\
                   &= - \log \left( \prod_{j=1}^{n_q} 
\frac{\exp(f(x_{qj}))}{\sum_{k=j}^{n_q} \exp(f(x_{qk}))}.\right)
\end{align}

Throughout this section, the candidate score list $x_q$ is assumed to be ordered according to the target permutation induced by $y_q$. In practice, we observe that $\mathcal{L}_{MLE}(x_q,y_q)$ grows with list length, which can destabilize training across queries with heterogeneous candidate sizes. We therefore employ a length-normalized variant:

\begin{align}
\overline{\mathcal{L}}_{MLE}(x_q,y_q)
    &= \frac{1}{n_q}\mathcal{L}_{MLE}(x_q,y_q)\\
    &= -\frac{1}{n_q}  \log \left( \prod_{j=1}^{n_q} 
\frac{\exp(f(x_{qj}))}{\sum_{k=j}^{n_q} \exp(f(x_{qk}))}\right).
\end{align}

Let $\mathcal{Q}_L$ and $\mathcal{Q}_U$ denote the labeled and unlabeled query sets, respectively. For q $\in$ $\mathcal{Q}_L$, we use $\overline{\mathcal{L}}_{MLE}(x_q,y_q)$. For q $\in$ $\mathcal{Q}_U$, to inject stable ordering signals from the upstream rankers, we add listwise distillation terms. Let $g^v$, $g^n$ denote the upstream visual ranker and text ranker, respectively. Acorddingly, let $\tilde{y}^v_q=\{g^v_\theta(x_{qj}) \}_{j=1}^{n_q}$ and $\tilde{y}^n_q=\{g^n_\theta(x_{qj}) \}_{j=1}^{n_q}$ denote the permutations sorted by their output scores. The final objective aggregates both supervision signals over the training set $\mathcal{Q}$:
$$
\mathcal{L}_{\text{final}}
= 
\overline{\mathcal{L}}_{\mathrm{MLE}}(x_q,y_q)+ \alpha\overline{\mathcal{L}}_{\mathrm{MLE}}^v(x_q,\tilde{y}^v_q)+\beta\overline{\mathcal{L}}_{\mathrm{MLE}}^n(x_q,\tilde{y}^n_q)
,
$$
where $\alpha$,$\beta$ balances label supervision and upstream supervisions.

\section{Experiments}
\subsection{Experimental Settings}
\subsubsection{Dataset}
To evaluate our approach, we adopt the public Qilin dataset~\cite{chen2025qilin} and Baidu’s DuRank dataset, which are both large-scale industrial datasets for heterogeneous ranking on The Red Book APPs and Baidu APPs, respectively. The detailed process of data is shown in Appendix \ref{sec:Data Processing}.
\begin{itemize}
    \item \textbf{Qilin:} Collected from The Red Book APPs, a large Chinese UGC platform, Qilin comprises more than 15,000 users’ app-level session contexts, 1,900,000 user-generated notes, and 5,000,000 images. It supports two primary tasks, i.e., search and recommendation. With its diverse multimodal content and rich user interactions, Qilin provides a challenging yet realistic benchmark for studying ranking over multimodal heterogeneous data.
    
    \item \textbf{DuRank:} We publish a high-quality dataset manually annotated by experts. This dataset is collected between May 2024 and March 2025, and contains more than 15,000 user queries, 300,000 retrieved results, and the corresponding human relevance judgments for each query-result pair. The detailed description of this dataset is shown in Appendix~\ref{sec:DuRank Dataset}. With its carefully curated annotations and coverage of real user search intents, this dataset serves as a valuable resource for evaluating ranking models in search scenarios. \textbf{The DuRank dataset will be publicly released after completing the required anonymization process and upon publication.}
\end{itemize}

\subsubsection{Metrics}
We evaluate our reranking methods using three classical ranking metrics for the main experiments, i.e., MRR, MAP, and NDCG. For the offline and online A/B testing, we additionally compare the ranking metrics, such as F1 and PNR, and the user experience metrics, such as IRQ, Next-day Retained User, and $\Delta GSB$. Due to space constraints, detailed descriptions of the Evaluation Metrics are provided in the Appendix~\ref{sec:Metrics}.

\subsubsection{Parameter settings}
The server is equipped with a Gigabit Ethernet card and utilizes multiple GPUs, including eight NVIDIA A100 and eight NVIDIA V100. FP32 is used to maximize performance for Bert-base-chinese and ERNIE (used in the online reranker), while FP16 and lora~\cite{hu2022lora} are used to accelerate the training for Qwen2-VL-2B and Qwen3-Reranker-4B. Other parameter settings are provided in the Appendix~\ref{sec:settings}.
\begin{table}[!t]
\centering
\resizebox{\linewidth}{!}{
\begin{tabular}{ccccc}
\toprule
 \multirowcell{2}{Method} & \multirowcell{2}{Labeled Size} &  \multicolumn{3}{c}{Qwen2-VL-2B} \\
\cmidrule(lr){3-5} 
 & &  MRR@10 & MAP@10 & NDCG \\
\midrule

 SFT &0\%-10\% data   & 0.6269 & 0.4699 & 0.4172 \\
 SFT &0\%-20\% data   & 0.6338 & 0.4913 & 0.4428 \\
 SFT &0\%-30\% data  & 0.6606 & 0.5016 & 0.4533 \\
 SFT & full data & 0.6664 & 0.5180 & 0.4693 \\ \hline
Only Upstream  & -   & 0.6172 & 0.4541 & 0.4038 \\
SMAR &random 30\%  & 0.6676 & 0.5125 & 0.4659 \\
SMAR &0\%-10\%   & 0.6702 & \textbf{0.5218} & \textbf{0.4756} \\
SMAR &0\%-20\%   & 0.6701 & 0.5184 & 0.4691 \\
SMAR &0\%-30\%   & \textbf{0.6738} & 0.5203 & 0.4722 \\
SMAR &30\%-70\% & 0.6715 & 0.5135 & 0.4670 \\
SMAR &70\%-100\% & 0.6582 & 0.4647 & 0.4507 \\
SMAR &full data  & 0.6687 & 0.5061 & 0.4600 \\ 
\cdashline{1-5}[3pt/3pt]
SMAR &0\%-20\% & \textbf{0.6701} & \textbf{0.5184} & \textbf{0.4691} \\
SMAR &20\%-40\% & 0.6677 & 0.5023 & 0.4564 \\
SMAR &40\%-60\% & 0.6676 & 0.5062 & 0.4556 \\
SMAR &60\%-80\% & 0.6684 & 0.4912 & 0.4433 \\
SMAR &80\%-100\% & 0.6419 & 0.4805 & 0.4307  \\ 

\bottomrule
\end{tabular}
}
\caption{Qilin: we test top-p sampling across budgets, where x\%–y\% is the slice of labeled instances by the upstream model’s ranking scores. We fine-tune SMAR on this selected data with human labels plus upstream supervision; the SFT baseline fine-tunes the Reranker only on the top-x\% data without single-modal ranker supervision. }
\label{tab:qilin-top-p-1}
\end{table}

\begin{table*}[t]
\centering
\begin{tabular}{cccccccc}
\toprule
 \multirowcell{2}{Method} & \multirowcell{2}{Labeled Size} & \multicolumn{3}{c}{Bert-base-chinese~\cite{devlin2019bert}}  & \multicolumn{3}{c}{Qwen3-Reranker-4B~\cite{qwen3embedding}} \\
\cmidrule(lr){3-5} \cmidrule(lr){6-8}
 & & MRR@1 & MRR@2  & NDCG & MRR@1 & MRR@2 & NDCG \\
\midrule
 SFT &1\% data   & 0.6097 & 0.7338 & 0.5762 & 0.6854 & 0.7858 & 0.6332 \\
 SFT &3\% data   & 0.6869 & 0.7873 & 0.6413 & 0.7097 & 0.8007 & 0.6595 \\
 SFT &5\% data   & 0.7078 & 0.7989 & 0.6538 & 0.7213 & 0.8071 & 0.6684 \\
 SFT &10\% data  & 0.7388 & 0.8196 & 0.6840 & 0.7269 & 0.8099 & 0.6710 \\
 SFT &20\% data  & 0.7366 & 0.8160 & 0.6817 & 0.7291 & 0.8142 & 0.6774 \\
 SFT &30\% data  & 0.7418 & 0.8235 & 0.6921 & 0.7377 & 0.8207 & 0.6841 \\
 SFT & full data & 0.7414 & 0.8218 & 0.6870 & 0.7537 & 0.8308 & 0.7008 \\
\cdashline{1-8}[3pt/3pt]
Only Upstream & -  & 0.7418 & 0.8248 & 0.6864 & 0.7276 & 0.8170 & 0.6751\\
SMAR &1\% data  & 0.7485 & 0.8285 & 0.6908 & 0.7347 & 0.8183 & 0.6778 \\
SMAR &3\% data  & 0.7474 & 0.8269 & 0.6881 & 0.7347 & 0.8198 & 0.6804 \\
SMAR &5\% data  & 0.7534 & 0.8317 & 0.6942 & 0.7340 & 0.8160 & 0.6801 \\
SMAR &10\% data  & 0.7526 & 0.8326 & 0.6947 & 0.7459 & 0.8224 & 0.6863 \\
SMAR &20\% data  & 0.7522 & 0.8338 & \textbf{0.6972} & 0.7410 & 0.8224 & 0.6852 \\
SMAR &30\% data  & \textbf{0.7586} & \textbf{0.8340} & 0.6952 &  0.7425 & 0.8256 & 0.6913 \\
SMAR & full data   & 0.7448 & 0.8285 & 0.6944 & \textbf{0.7559} & \textbf{0.8341} & \textbf{0.7014} \\
\bottomrule
\end{tabular}
\caption{DuRank: compares retrieval performance across data sizes. The SFT baseline fine-tunes the reranker on a random x\% of queries without single-modal ranker supervision. x\% of the dataset is treated at the query level, rather than item level.}
\label{tab:DuRank dataset}
\end{table*}

\subsection{Main Results}

\subsubsection{Qilin Dataset}
To simulate industrial candidate generation, we train two upstream retrieval models on the Qilin dataset by modality: BERT-base-Chinese for text and Qwen-VL-2B for vision-dominant items. Each retriever independently returns a ranked list with relevance scores. The union of these lists forms the multimodal candidate pool and preserves the original upstream ordering as a prior. Under a constrained human and time budget, we leverage the scores generated by upstream single-modality rankers on their respective candidate pools and distill their modality-specific ranking preferences into a unified cross-modality ranking model. The goal is to achieve near full-supervision performance while minimizing the need for expensive and time-consuming annotations.

Experimental results on the Qilin Dataset are shown in Table~\ref{tab:qilin-top-p-1}. Using scores from upstream single-modality models for supervision consistently improves the performance of all models trained on the annotated and cross-modality dataset. Moreover, when labeled data are limited, comparing models trained on different quantile segments reveals that annotating only the top-ranked portion of each queue yields better performance than randomly annotating samples or labeling those from the lower-ranked tail. Remarkably, using only the top 10\% of annotated data, the proposed SMAR model surpasses the vanilla model trained on the full annotation set without single-modality supervision across all evaluation metrics. 




\begin{table}[t]
\centering
\resizebox{\linewidth}{!}{
\begin{tabular}{cccccc}
\toprule
Method & $T_{round}$ & Labeled Size &  F1 & NDCG@4 & pnr\\
\midrule
Baseline & - & - & 0.7258 & 0.8766 & 1.885 \\
Vanilla & - & 100\% & 0.7211 & 0.8788 & \textbf{2.1527} \\
SMAR-binary search & 1 & 42.09\% & 0.7134 & 0.8803 & 2.0619\\
SMAR-binary search & 2 & 58.62\% & \textbf{0.7269} & 0.8791 & 2.1492\\
SMAR & - & 100\%  & 0.7266 & \textbf{0.8842} & 2.1342 \\
\bottomrule
\end{tabular}
}
\caption{Offline Metrics of Binary Search for Iso-Label Anchors on Baidu APPs. }
\label{tab: binary search}
\end{table}

\subsubsection{DuRank Dataset}
In the DuRank dataset,  upstream signals are derived from independently developed in-house ranking systems. Each query is associated with an average of 4.4 candidate items, which is insufficient for applying the top-p sampling strategy. Therefore, x\% of the dataset is treated at the query level. We require at least two natural and two video candidates per query, balance the top-ranked items across modalities as evenly as possible, and discard queries whose maximum relevance score is below 0.8. The BERT-base-chinese is trained with 30 epochs. Due to time constraints, qwen3-reranker is trained with only 15 epochs. 

The results of the same experiments on DuRank are shown in the Table~\ref{tab:DuRank dataset}. Experimental results demonstrate that SMAR remains effective across different model architectures, including the decoder-only Qwen3-Reranker and the encoder-based BERT-Base-Chinese, suggesting the general applicability of our approach. Notably, BERT-Base-Chinese, trained with only 30\% of the labeled data, already surpasses its counterpart trained on the full annotation set. In contrast, Qwen3-Reranker exhibits performance gains under single-modality supervision but still falls short of the full annotated data.

\subsubsection{Binary Search for Iso-label Anchors}


\begin{table}[!t]
\centering
\resizebox{\linewidth}{!}{
\begin{tabular}{cccccc}
\toprule
 Modality & Method & MRR@10 & MAP@10 & NDCG \\
\midrule
text &  - & 0.5530 & 0.3969 & 0.3416 \\
visual & - & 0.5776 & 0.4140 & 0.3623 \\
multimodal & MLP & 0.6021 & 0.4652 & 0.4089 \\
multimodal & self-attention~\cite{pobrotyn2020context} & 0.6168 & 0.4785 & 0.4251 \\
multimodal(SMAR) & cross-attention(w/o hybrid fusion) & 0.6633 & 0.4996 & 0.4579 \\
multimodal(SMAR) & cross-attention & \textbf{0.6664} & \textbf{0.5180} & \textbf{0.4693} \\

\bottomrule
\end{tabular}
}
\caption{Whole-page reranker with different attention modules on Qilin Dataset. All methods use the hybrid fusion. Training excludes the upstream-ranker supervision.}
\label{tab:Cross-Atten}
\end{table}

The results of binary search for iso-label anchors are shown in Table~\ref{tab: binary search}. ~\footnote{The iso-label anchor is suitable for discrete bucketed features. Since the model outputs in the Qilin dataset are not bucketed, we conducted this experiment only on DuRank.} 
When only two rounds of binary search are conducted, SMAR achieves superior performance on both F1 and NDCG while using merely 58.62\% of the original dataset. Its PNR score also approaches that of the fully human-annotated supervised model, with a gap of less than 0.01.


\subsubsection{Cross-Attention Mechanism}
 We train the whole-page reranker on the annotated training set to assess the effectiveness of the proposed cross-attention mechanism on the Qilin dataset. The results are reported in Table ~\ref{tab:Cross-Atten}. The rows corresponding to the text and visual modalities represent the performance achieved when models are trained using a single modality on the multimodal ranking task. The experimental results show that the cross-attention architecture incorporating user features significantly outperforms other methods, and the combination of early and late fusion for item representation further amplifies this advantage.

\subsection{Online A/B Testing}

\subsubsection{Settings}
Except for the online model, which uses Baidu ERNIE~\cite{ernie2025technicalreport}, all other whole-page rerankers in this paper are trained with cross-attention and a listwise objective.
To fully evaluate the reranking performance of SMAR, we adopt pairwise supervision to distill the upstream single-modal ability into our multimodal relevance model. For each query, results with higher upstream scores are treated as positive samples, while those with lower scores are treated as negative samples. The loss function is 
\begin{align*}
    loss &=loss_{pointwise}+ \alpha \cdot loss_{label\ pairwsie}+ \beta \cdot loss_{upstream\ pairwsie}\\
         &=loss(pred,label) + \alpha \cdot max\left( margin_1-(s_{pos}-s_{neg}),0\right)\\
         &+ \beta \cdot max\left( margin_2-(s_{pos}-s_{neg}),0\right).
\end{align*}
The hyperparameters were set as follows: $margin_1=0.1$, $margin_2=0.1$, $\alpha=0.5$, and $\beta=0.2$, while $s_{pos}$ and $s_{neg}$ denote the predicted relevance of the positive and negative samples. During the online A/B test, we performed a seven-day experiment (from 2025-09-16 to 2025-10-01) to compare it with our online relevance reranker in a real-world search scenario.




\subsubsection{Results}
\begin{figure}[t]
\centering
\includegraphics[width=\columnwidth]{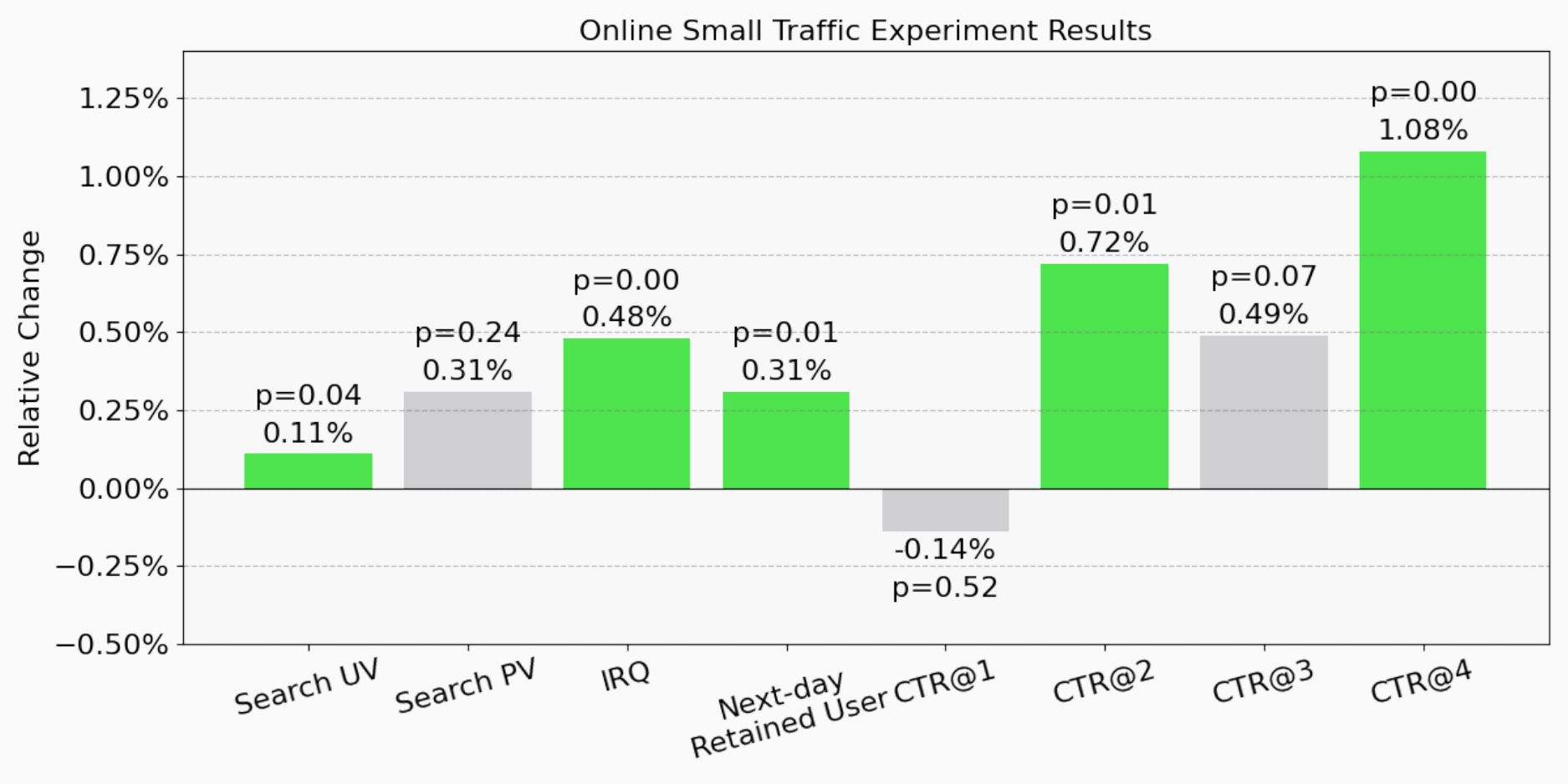}
\caption{\textbf{Online A/B test.} Bars report the relative change (\%) of each metric for the treatment vs.\ baseline. The $y$-axis shows a zero-effect baseline. Each bar is annotated with its effect size and $p$-value. Color encodes statistical significance: \emph{green} denotes a significant improvement ($p\le 0.05$ and $\Delta>0$) and \emph{gray} denotes a non-significant effect ($p>0.05$).}
\Description{The figure shows the results of an online A/B test comparing the treatment model with the baseline across multiple metrics. The vertical axis represents the relative change percentage, with a zero-effect baseline. Bars for Search UV, Search PV, IRQ, Next-day Retained User, and click-through rates (CTR@2–4) show positive improvements, while CTR@1 slightly decreases. Green bars indicate statistically significant gains (e.g., +0.48\% for IRQ and +1.08\% for CTR@4), gray bars denote non-significant effects, and no red bars appear. Overall, the treatment demonstrates consistent and significant performance gains.}

\label{fig: online Experiment}
\end{figure}

All effects in Figure \ref{fig: online Experiment} are reported as relative percentage changes vs. baseline. The treatment increases scale indicators: Search UV (unique users who used search) rises by 0.11\% (p=0.04), indicating a significant gain; Search PV (search page views) increases by 0.31\% (p=0.24); and IRQ—the total quantity of results issued, a proxy for search distribution and contributed information—shows a significant 0.48\% lift ($p\approx0.00$). Together with a significant improvement in next-day retention (0.31\%, p=0.01), these results suggest greater user engagement and perceived relevance. On relevance at different ranks, we analyze the cause of the slight decrease in CTR@1. Through extensive case studies, we observe that the treatment model tends to rank generative answer cards produced by LLMs at the top position. These cards are typically self-contained and satisfy user intent without requiring click-through, which explains the (non-significant) reduction in CTR@1. The treatment boosts mid/tail click propensity, indicating higher-quality ranking beyond the first position and deeper exploration. In addition to CTR@k, aggregated CTR over the top-4 and top-8 increases by 0.16\% and 0.25\%, respectively. Finally, we observe $\Delta\mathrm{GSB}=1.58\%$ on randomly sampled queries with 190 effective expert judgements and 0.50\% on long-tail queries with 193 effective expert judgements, demonstrating that SMAR outperforms the latest online model under expert human evaluation.




\section{Related Work}
\subsection{Learning to Rank}
Learning to Rank (LTR) has been extensively studied in search and recommendation systems, and existing methods generally fall into three paradigms: pointwise, pairwise, and listwise. Pointwise approaches \cite{guo2017deepfm} treat ranking as a regression or classification task, predicting each item’s relevance independently. Pairwise methods \cite{wan2022cross,burges2006learning} optimize relative preferences by training on positive–negative item pairs, improving discriminative capability. Listwise approaches further consider the entire ranking list, optimizing list-level metrics and modeling item dependencies more effectively. Attention mechanism, as a listwise approach, is designed for explicitly capturing inter-item relationships. Pobrotyn et al. \cite{pobrotyn2020context} introduced a context-aware listwise model using self-attention to model pairwise interactions within a candidate list. However, such models often overlook personalization, assuming uniform user preferences. 
To address this, we propose Cross-Attention-Rank, which incorporates the user features via cross-attention, aligning user embeddings with item features for personalized ranking.

\subsection{Multimodal Ranking}
Recent work in multimodal ranking explores integrating visual and textual signals through Multimodal Large Language Models (MLLMs). While these models aim for unified multimodal representations, fine-tuning often favors textual modalities, weakening visual contributions. To mitigate this bias, NoteLLM-2 \cite{zhang2025notellm} introduces multimodal in-context learning (mICL) with late fusion, separating visual and textual prompts and applying a gating mechanism to effectively preserve visual information, leading to superior performance in multimodal recommendation. Similarly, the Joint Relevance Estimation (JRE) framework \cite{zhang2018relevance} fuses screenshots, HTML, and text of search result pages using inter- and intra-modality attention to balance heterogeneous signals. However, JRE primarily addresses intra-page reranking and does not resolve the scoring misalignment that arises when jointly ranking heterogeneous modality items lacking shared objectives.

\section{Conclusion}

We introduced SMAR, a whole-page reranking framework that aligns cross-modal relevance by distilling signals from strong single-modal rankers. Rather than relying on expensive page-level labels, SMAR uses Top-P and Iso-label anchors to focus limited annotations, preserving intra-modality order while improving cross-modal ranking ability. Experiments on Qilin and DuRank show that SMAR cuts annotation costs by ~70–90\% and delivers consistent gains in NDCG, CTR, and downstream engagement in offline evaluations and online A/B tests. These results highlight that coupling single-modal expertise with budget-aware alignment is a practical and scalable path to higher-quality SERP. Future work includes adaptive distillation that responds to intent and distribution shifts, broader modality coverage (e.g., code, agents), and deeper integration with LLM-generated candidates to enhance the user experience.

\bibliographystyle{ACM-Reference-Format}
\bibliography{MultimodalRanking}

\appendix
\section{Metrics}
\label{sec:Metrics}
\subsection{Offline Metrics}
\begin{itemize}
    \item \textbf{MRR$@$k} (Mean Reciprocal Rank) \\
    MRR$@$k measures how well the model ranks the \emph{first relevant result} at a high position. Higher MRR indicates that users can find relevant items earlier. 
    For a set of queries $\mathcal{Q}$, let $\text{rank}_i$ denote the rank position of the first relevant item for query $q_i$. Then
    \[
    \text{MRR@}k = \frac{1}{|\mathcal{Q}|} \sum_{i=1}^{|\mathcal{Q}|} \frac{\mathbb{I}(\text{rank}_i \leq k)}{\text{rank}_i},
    \]
    where $\mathbb{I}(\cdot)$ is the indicator function.

    \item \textbf{MAP$@$k} (Mean Average Precision) \\
    MAP$@$k measures the model's ability to rank \emph{all relevant items} higher, not just the first one. Higher MAP indicates better ranking consistency across relevant results. For query $q_i$, 
    \[
    P_i@j = \frac{1}{j}\sum_{t=1}^{j}\text{rel}_i(t).\]
    where,
    \[\text{rel}_i(t) =
    \begin{cases}
    1, & \text{if item at rank $t$ is relevant}, \\
    0, & \text{otherwise}.
    \end{cases}\]
    The average precision at cutoff $k$ is
    \[
    AP_i@k = \frac{1}{R_i} \sum_{j=1}^{k} P_i(j)\cdot \text{rel}_i(j),
    \]
    where $R_i = \sum_{j=1}^{k} \text{rel}_i(j)$. The mean average precision is
    \[
    \text{MAP@}k = \frac{1}{|\mathcal{Q}|}\sum_{i=1}^{|\mathcal{Q}|} AP_i@k.
    \]

    \item \textbf{NDCG} (Normalized Discounted Cumulative Gain) \\
    NDCG measures the ability of the model to rank highly relevant items near the top, with emphasis on both position and graded relevance.  
    For query $q_i$, the discounted cumulative gain is:
    \[
    DCG_i = \sum_{j=1}^{N} \frac{2^{\text{rel}_i(j)} - 1}{\log_2(j+1)}
    \]
    where N is the number of retrieved items for query $q_i$ and $\text{rel}_i(j)$ is the graded relevance of the item at rank $j$. The ideal DCG (IDCG) is computed by sorting items by true relevance. Then
    \[
    \text{NDCG} = \frac{1}{|\mathcal{Q}|} \sum_{i=1}^{|\mathcal{Q}|} \frac{DCG_i}{IDCG_i}
    \]

    \item \textbf{F1} (Harmonic Mean of Precision and Recall) \\
    F1 evaluates the balance between precision and recall without relying on a cutoff. For query $q_i$, let $\mathcal{S}_i$ be the set of items returned by the system (e.g., the full ranked list or those above a fixed decision threshold), and let $\mathcal{R}_i$ be the set of ground-truth relevant items with $r_i = |\mathcal{R}_i|$. Define
    \[
    P_i = 
    \begin{cases}
    \frac{|\mathcal{S}_i \cap \mathcal{R}_i|}{|\mathcal{S}_i|}, & |\mathcal{S}_i|>0,\\[2pt]
    0, & \text{otherwise},
    \end{cases}
    \qquad
    R_i = \frac{|\mathcal{S}_i \cap \mathcal{R}_i|}{r_i}.
    \]
    The per-query F1 score is
    \[
    F1_i = 
    \begin{cases}
    \displaystyle \frac{2\,P_i R_i}{P_i + R_i}, & P_i + R_i>0,\\[8pt]
    0, & \text{otherwise}.
    \end{cases}
    \]
    We report the macro-average over queries with at least one relevant item:
    \[
    \text{F1} = \frac{1}{|\mathcal{Q}'|}\sum_{i \in \mathcal{Q}'} F1_i, \quad \mathcal{Q}'=\{\,i \mid r_i>0\,\}.
    \]

    \item \textbf{PNR} (Positive–Negative Ratio) \\
    PNR quantifies pairwise order consistency between model scores and ground-truth relevance. For query $q_i$ with candidates indexed by $a\in\{1,\ldots,n_i\}$, let $y_{i,a}$ denote the ground-truth relevance (larger is more relevant) and $s_{i,a}$ the model score. Define the counts of concordant and discordant labeled pairs as

    \begin{align}
        C_i&=\sum_{a=1}^{n_i}\sum_{b=1}^{n_i}\mathbb{I}\!\big(y_{i,a}>y_{i,b}\big)\,\mathbb{I}\!\big(s_{i,a}>s_{i,b}\big),\\
        D_i&=\sum_{a=1}^{n_i}\sum_{b=1}^{n_i}\mathbb{I}\!\big(y_{i,a}>y_{i,b}\big)\,\mathbb{I}\!\big(s_{i,a}<s_{i,b}\big).
    \end{align}

    Label ties are excluded by construction; score ties are ignored. Then
    \[
    \text{PNR}_i=\frac{C_i}{D_i},
    \]
    and the dataset-level metric is the macro-average
    \[
    \text{PNR}=\frac{1}{|\mathcal{Q}|}\sum_{i=1}^{|\mathcal{Q}|}\text{PNR}_i.
    \]
    Higher PNR indicates fewer pairwise inversions and stronger alignment with ground truth.

\end{itemize}

\subsection{Online Metrics}
\textbf{Search UV} counts the number of unique users who interacted with the search surface.
\textbf{Search PV} counts the number of search page views.
\textbf{Issued Result Quantity (IRQ)} is the number of results delivered to users and subsequently viewed. It is a key indicator of search scale and, to some extent, quantifies the amount of information contributed by the search engine.
\textbf{Next-day Retained User} refers to the fraction of users who return the following day.
For click metrics, \textbf{CTR@k} denotes the position-specific click-through rate at rank position $k$ (clicks on position $k$ divided by impressions of position $k$), for $k\in\{1,2,3,4\}$.
Relative to the online baseline, we obtain $N$ expert pairwise judgments. Letting \#good and \#bad denote counts favoring our model vs.\ baseline, we define
\[
\Delta\text{GSB}=\frac{\#\text{good}-\#\text{bad}}{2\times(\#\text{good}+\#\text{bad}+\#\text{same})}\,.
\]
Positive $\Delta\text{GSB}$ indicates a net human preference for our model. 

\section{DuRank Dataset}
\label{sec:DuRank Dataset}


The DuRank dataset was constructed from real user interactions on the Baidu App. It contains query data and high-quality annotations obtained through expert labeling. The dataset primarily consists of search results retrieved from Baidu’s natural ranking channel and video ranking channel, with more than 200,000 video results and 40,000 natural results included. Each sample contains a query, the corresponding retrieved results with titles, summaries, categorizations, upstream model scores, and human-annotated relevance, etc. The DuRank dataset comprises single-modality and multi-modality corpora. The single-modality corpus provides only a training set, where all results for a given query come from the same source, either natural search results or video results, and each query is paired with at least two results. The multi-modality corpus includes both training and test sets, and each query is associated with at least two results drawn from different modalities. Expert-annotated relevance is provided on a five-point scale ranging from 0 to 4, indicating the degree to which an item satisfies the user query. 
A score of 0.0 denotes that the item fails to meet the query at all, while a score of 4.0 denotes that the item fully satisfies the query.
Owing to its high-quality human annotations, the Durank dataset enables search engines to better capture the relevance between user queries and retrieved results, thereby providing a valuable dataset for multimodal ranking research.

\section{Parameter settings}
\label{sec:settings}

In the Qilin experiments, we train for 30 epochs with a batch size of 1 at the query level, where each query contains multiple candidate items whose total number corresponds to the actual model input batch size. The maximum sequence length is set to 512, which is sufficient for nearly all items, and the learning rate is $1\times10^{-5}$.

In the DuRank experiments, for the Qwen3-Reranker model, we train for 15 epochs with a learning rate of $2\times10^{-6}$ and a batch size of 2, applying LoRA fine-tuning to both attention and MLP layers within the attention blocks. For the BERT-Base-Chinese model, we train for 30 epochs with a batch size of 32, a maximum sequence length of 512, and a learning rate of $1\times10^{-5}$.

\section{Data Processing}
\label{sec:Data Processing}
\begin{figure}[h]
\centering
\includegraphics[width=0.9\linewidth]{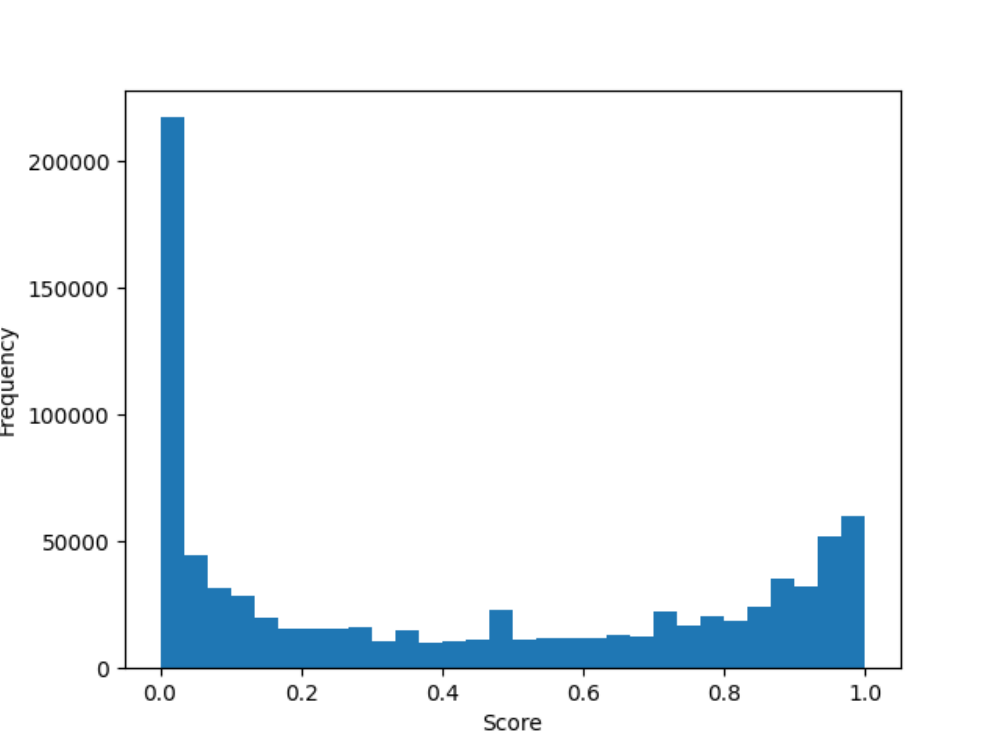}
\caption{Distribution of QwenReranker scores for candidate items in the Qilin dataset. Scores are used to distinguish text-driven and image-driven clicks.}
\label{fig:qwen_score_distribution}
\Description{The figure shows a histogram of QwenReranker scores for candidate items in the Qilin dataset. 
The horizontal axis represents the score values from 0 to 1, and the vertical axis represents frequency. 
The distribution exhibits a concentration of items near 0 and another increase near 1. These scores are used to distinguish between text-driven and image-driven clicks.}
\end{figure}

\begin{algorithm}[!t]
\caption{Binary search for iso-label anchors}
\label{alg:hybrid_iso_label}
\begin{algorithmic}[1]
\Require Two modality queues $Q_1$ (video) and $Q_2$ (natural), sorted by upstream scores; maximum rounds $T_{\text{rounds}}$
\Ensure Cross-modality sequence $\mathcal{S}$ containing \texttt{tie}, \texttt{virtual\_tie}, and \texttt{single} segments.

\State Initialize $\mathcal{S} \gets [\,]$, and search index $i \gets 0$ in $Q_1$
\For{$i = 1$ to $|Q_1|$}
    \State $a \gets Q_1[i]$
    \State $k \gets$ \textbf{BinarySearch} in $Q_2$ for item with $Q_2[k].\text{label} = a.\text{label}$
    \If{$k$ found}
        \State Append \texttt{tie}$(a, Q_2[k])$ to $\mathcal{S}$
        \State $T_{\text{rounds}}+=1$
    \ElsIf{there exist higher and lower labels around $a.\text{label}$ in $Q_2$}
        \State Append \texttt{virtual\_tie}$(a, a^{(\text{virtual})})$ to $\mathcal{S}$
        \State $T_{\text{rounds}}+=1$
    \ElsIf{all labels in $Q_2[j:] > a.label$}
        \State Append $concatenate(Q_2[j:],a)$ as \texttt{single} to $\mathcal{S}$
    \ElsIf{all labels in $Q_2[j:] < a.label$}
        \State Append $concatenate(a,Q_2[j:])$ as \texttt{single} to $\mathcal{S}$  
    \EndIf
    \If{rounds exceed $T_{\text{rounds}}$}
        \State Append remaining items as \texttt{single} segments; \textbf{break}
    \EndIf
\EndFor
\State \Return $\mathcal{S}$
\end{algorithmic}
\end{algorithm}

To facilitate both single-modality and mixed-modality ranking experiments, we reorganize the original Qilin dataset into dedicated training and test sets, ensuring that evaluation metrics reflect realistic ranking scenarios. In particular, we enforce that, for each query in the test set, the number of non-clicked items exceeds the number of clicked items, ideally by a factor of three. This design prevents artificially inflated MRR scores that could arise if all candidate items under a query are clicked.

For single-modality data, we split the training set according to text and image modalities. For textual data, each candidate document for a given query is first scored by QwenReranker using the concatenated query-document input. Among the clicked items, those with a Qwen3 reranker similarity score below 0.1 are treated as image-driven clicks and are excluded from the textual training set. The threshold 0.1 is set manually after careful inspection. Conversely, clicked items with a score above 0.1 are retained as positive samples. Non-clicked items, regardless of modality, are included as negative samples, forming a balanced training set for text ranking. For image-based ranking, since user clicks are naturally driven by both images and titles, we retain the original dataset without further filtering. These two datasets are independent for training models of different modalities.

The test set is processed in a similar fashion. For text ranking, clicked items with a low Qwen score are considered image-driven and their click label is set to 0, while clicked items with a high score retain a click label of 1. Non-clicked items are consistently labeled as 0. Image-based test data remain unchanged, as user interactions already reflect the joint effect of images and titles. For listwise objection, all candidate items are further sorted according to Wilson-smoothed click-through rates.

For mixed-modality ranking, we distinguish the modality of candidate items for each query. Clicked items with low textual relevance are treated as image-driven, while high textual scores are assumed to text-driven, though potential image contributions are ignored. The ranking objective in this setting is to ensure that clicked items are placed above non-clicked ones, allowing the model to learn to integrate signals across modalities effectively.

Figure~\ref{fig:qwen_score_distribution} presents the distribution of QwenReranker scores across candidate items, illustrating the separation between text- and image-driven clicks and providing guidance for subsequent model training and evaluation.

For the text modality, we concatenate each item's title and content as its textual representation. For the visual modality, the image title is incorporated into the visual-language model (VLM) prompt, while the image itself is used as input to the ViT component of Qwen-VLM.

\begin{algorithm}[!t]
\caption{Pair construction based on iso-label anchors}
\label{alg:pair_construction}
\begin{algorithmic}[1]
\Require Aligned sequence $\mathcal{S}$ with optional labels or aurora scores
\Ensure Pair set $\mathcal{P}$ for pairwise supervision, and pointwise set $\mathcal{P}^0$

\State $\mathcal{P} \gets \emptyset$, $\mathcal{P}^0 \gets \emptyset$

\ForAll{pairs $(x, y)$ within or across queues in $\mathcal{S}$}
    \If{$x$ and $y$ have labels}
        \If{$x.\text{label} > y.\text{label}$}
            \State $\mathcal{P} \gets \mathcal{P} \cup \{(x, y)\}$ 
        \Else
            \State $\mathcal{P} \gets \mathcal{P} \cup \{(y, x)\}$ 
        \EndIf
    \ElsIf{labels unavailable and $x, y$ from same queue}
        \If{$x.\text{aurora} > y.\text{aurora}$}
            \State $\mathcal{P} \gets \mathcal{P} \cup \{(x, y)\}$ 
        \Else
            \State $\mathcal{P} \gets \mathcal{P} \cup \{(y, x)\}$ 
        \EndIf
    \EndIf
\EndFor

\ForAll{item $x$ with known label}
    \State $\mathcal{P}^0 \gets \mathcal{P}^0 \cup \{(x, x)\}$ \Comment{pointwise supervision}
\EndFor

\State \Return $(\mathcal{P}, \mathcal{P}^0)$
\end{algorithmic}
\end{algorithm}

\subsection{Feature Selection via Entropy-Based Sorting}

To mitigate the high cost of manual annotation while preserving discriminative features, we employ entropy-based sorting for numerical feature selection ~\cite{brown2012conditional}. The entropy H(X) of each feature X is calculated to quantify its information content, with features ranked in descending order of entropy. This ensures that features with higher information gain (e.g., aurora scoring from upstream models) are prioritized over redundant ones (e.g., howto). The final selected numerical features include QUERY\_SEARCH, SCS\_ERNIE, aurora, CLICK\_SKIP\_RATIO, and others as listed in Table 1, while categorical features like is\_wenda are encoded as binary indicators.

For the cross-modality ranking model in Qilin, we align aurora score sequences with human-annotated 5-point labels (0-4) to minimize the discrepancy between predicted and true rankings. The dataset is partitioned into training (70\%), validation (15\%), and test (15\%) sets, ensuring sufficient candidate diversity to compute robust metrics like MRR and NDCG.

\section{Binary Search for Iso-Label Anchors}
\label{sec: Binary Search for Iso-Label Anchors}

Algorithm~\ref{alg:hybrid_iso_label} employs a binary search strategy to locate items with identical relevance labels across different modalities. Then in algorithm~\ref{alg:pair_construction}, the limited amount of labeled data serves as pointwise supervision, guiding the cross-modality ranking model to predict accurate prior relevance scores for items. For unlabeled data, items are distinguished according to the modality-specific scores produced by the respective upstream single-modality models.

\end{document}